\documentclass[10pt, a4paper,twocolumn]{article}
\def\imscale{.9} %twocolumn

\usepackage[T1]{fontenc}
\usepackage[utf8]{inputenc}
\usepackage{hyperref}
\hypersetup{colorlinks=true,citecolor=blue,linkcolor=blue,urlcolor=blue}
\usepackage{natbib}
\usepackage[top=1in, bottom=1.25in, left=.75in, right=.75in]{geometry}
\usepackage{amssymb}
\usepackage{enumerate}
\usepackage{graphicx} %For figures
\usepackage{amsmath} %For matrix displays
\usepackage[english]{babel}

\usepackage{relsize}
\usepackage{units} % set units and nicefrac (nicefrac package included)
\usepackage{bm} %bold math
\usepackage{setspace}
\usepackage{subcaption}
\usepackage{xfrac}
\usepackage{gensymb}
\usepackage[mathlines]{lineno} %line numbering

\numberwithin{equation}{section} %sets equation numbers <chapter>.<section>.<index>
\numberwithin{figure}{section} %sets figure numbers <chapter>.<section>.<index>

\newcommand{\pdiff}[2]{\frac{\partial #1}{\partial #2}}
\newcommand{\ddiff}[2]{\frac{\mathrm d #1}{\mathrm d #2}}
\newcommand{\dddiff}[2]{\frac{\mathrm d^2 #1}{\mathrm d #2^2}}

\newcommand{\brac}[1]{ \left(  #1 \right) }

\newcommand{\diff}[1]{\left[ #1 \right]}
\newcommand{\bdiff}[1]{\big[ #1 \big]}
\newcommand{\diffx}[1]{\diff{#1}\_L\^R}

\newcommand{\diffpm}[1]{\diff{#1}\_-\^+}
\newcommand{\bdiffpm}[1]{\bdiff{#1}\_-\^+}

\newcommand{\diffk}[1]{\diff{#1}\g^\ell}
\newcommand{\an}[1]{\left\langle #1 \right\rangle}

\newcommand{\ie}{{\textit{i.e.}}}
\newcommand{\egg}{\textit{e.g.}}

\let\underscore\_
\renewcommand{\_}[1]{_\mathrm{#1} }

\let\texthat\^
\renewcommand{\^}[1]{^\mathrm{#1} }

\newcommand{\of}[1]{\mbox {\smaller$\brac{#1}$}}
\newcommand{\inv}{^{\raisebox{.2ex}{$\scriptscriptstyle-1$}}}

\newcommand{\wt}{\tilde}
\newcommand{\wh}{\hat}
\newcommand{\ol}{\overline}
\newcommand{\mr}{\mathrm}
\newcommand{\mc}{\mathcal}
\renewcommand{\l}{\_{\ell}}
\newcommand{\g}{\_g}
\renewcommand{\k}{_\phaseindex}
\newcommand{\phaseindex}{k}
\renewcommand{\a}{a}
\newcommand{\al}{\a\l}
\newcommand{\ag}{\a\g}
\newcommand{\ak}{\a\k}
\newcommand{\dx}{\Delta x}
\newcommand{\dt}{\Delta t}

\newcommand{\h}{\frac12}

\newcommand{\nn}{^{\mr{new}}}

\newcommand{\n}{n}
\newcommand{\const}{\mr{const}}
\newcommand{\Em}{\mbox{\sc{e}-}}

\newcommand{\Ur}{U\_r}

\newcommand{\Ukr}{U_{\phaseindex,\mr r}}
\newcommand{\Ulr}{U\_{\ell,r}}
\newcommand{\Ugr}{U\_{g, r}}
\newcommand{\ukr}{u\_{\ell,r}}
\newcommand{\ukpr}{\up_{\phaseindex,\rm r}}
\newcommand{\ulpr}{\up_{\ell,\rm r}}
\newcommand{\upr}{\up\_r}

\newcommand{\area}{\mc A}
\newcommand{\Qm}{\mc Q}
\newcommand{\A}{A}
\newcommand{\Ag}{\A\g}
\newcommand{\Al}{\A\l}
\newcommand{\Ak}{\A\k}

\newcommand{\Alnil}{\A_{\ell,0}}

\newcommand{\br}{\brac}

\newcommand{\dAl}{\A_{\ell,\x}}
\newcommand{\ddAl}{\A_{\ell,\x\x}}

\newcommand{\dA}{\A_{\x}}

\newcommand{\dU}{U_\x}

\newcommand{\f}{\wh f}
\newcommand{\F}{\wh F}
\newcommand{\df}{\f_{\x}}
\newcommand{\ddf}{\f_{\x\x}}
\newcommand{\dfnil}{\f_{\x,0}}

\newcommand{\up}{\wh u}
\newcommand{\ap}{\wh a}
\newcommand{\eps}{\varepsilon}

\newcommand{\epsp}{\wh\eps}

\newcommand{\mx}{w_x}
\newcommand{\my}{w_y}
\newcommand{\Drhogcos}{\my \dHdAl }
\newcommand{\x}{\xi}
\newcommand{\xshock}{\x_\pm}
\renewcommand{\b}{b}

\newcommand{\Gm}{m}
\newcommand{\Alnull}{\A_{\ell,0}}
\newcommand{\Q}{Q}
\newcommand{\Qr}{\Q\_r}
\newcommand{\Qkr}{\Q_{\phaseindex,\mr r}}

\newcommand{\Ql}{\Q\l}
\newcommand{\Qg}{\Q\g}
\newcommand{\q}{q}
\newcommand{\ql}{\q\l}
\newcommand{\qg}{\q\g}
\newcommand{\qk}{\q\k}
\newcommand{\qlr}{\q\_{\ell,r}}
\renewcommand{\j}{j}
\newcommand{\jr}{\j\_r}
\newcommand{\Jr}{J\_r}
\newcommand{\N}{\Jr'}

\newcommand{\dHdAl}{\mc H'}
\newcommand{\ddHdAl}{\mc H''}

\newcommand{\mcS}{\mc S}
\renewcommand{\S}{S}
\newcommand{\diffU}{\mcS_{\Ql}-\mcS_{\Qg}}
\newcommand{\diffA}{ \mcS_{\Al}}

\newcommand{\coeff}{\mc R}
\newcommand{\cftwo}{\coeff_{2}}
\newcommand{\cfone}{\coeff_{1}}
\newcommand{\cfnil}{\coeff_0}

\newcommand{\USL}{\ol\ql/\area}
\newcommand{\USG}{\ol\qg/\area}

\newcommand{\bw}{\bm v}
\newcommand{\bwp}{\bm w}%{\bw\^p}
\newcommand{\bwpt}{\widetilde \bwp}%{\wt\bw\^p}
\newcommand{\Jac}{\mathbb  A}
\newcommand{\Jach}{\mathbb  A\_h}
\newcommand{\Jacu}{\mathbb  A\_u}
\newcommand{\bff}{\bm f}
\newcommand{\bfu}{\bm f\_u}
\newcommand{\bfh}{\bm f\_h}
\newcommand{\ws}{ z}%{\bw\^p}
\newcommand{\bs}{\bm s}
\newcommand{\Hofal}{\mc H}
\newcommand{\dHofal}{\Hofal'}
\newcommand{\LL}{\mathbb  L}
\newcommand{\Lamb}{\mathbb \Lambda}

\newcommand{\Jph}{_{j+\frac12}}
\newcommand{\Jmh}{_{j-\frac12}}

\newcommand{\Jm}{_{j-1}}
\newcommand{\CFL}{}

\newcommand{\m}{^*}

\newcommand{\sign}{\mr{sign}}

\title{ The Stability of Roll-Waves in Two-Phase Pipe Flow }

\author{
\textsc{A.H.~Akselsen}\\[1ex]
\normalsize Norwegian University of Science and Technology, \\
\normalsize Department of Energy and Process Engineering, Kolbj{\o}rn Hejes v.\ 1B, 7491 Trondheim, Norway. \\ 
\normalsize \href{mailto:andreas.h.akselsen@ntnu.no}{andreas.h.akselsen@ntnu.no}
}

\date{ }

\begin{document}
%\linenumbers

\maketitle

\begin{abstract}
Roll-wave trains constitutes a well-known two-phase flow regime in pipes. There exists a one-parameter family of steady roll-wave train solutions, provided the flow conditions are within the roll-wave range. This means that wave train solutions can be constructed from out of a wide range of wavelengths. That band of wavelengths which will be observed in nature is however fairly narrow. The wavelength distribution is believed to be related to wave train stability and the flow disturbances.

Steady roll-wave train solutions are in this article  subjected to a linear stability analysis. Comparisons are made with predictions from direct numerical Roe scheme simulations. Good agreement is observed; after an initial stage of wave coalescence, simulated wavelengths are distributed among the shorter of those wavelengths which are predicted linearly stable. Also the observed disturbance frequency and rate of decay agrees with the analysis predictions. Finally, the stability of pressure driven gas-liquid trains is compared to that of gravity driven free surface trains.
\end{abstract}

\section{Introduction}
Roll-waves trains consist of a series of exponentially profiled wave structures, often called `bores', connected by hydraulic jumps. 
Thomas~\cite{Thomas_roll_waves} was amongst the first to publish analytical expressions for the roll-wave profiles of channel flows using a moving belt analogy. He also provided insight into the conditional nature of roll-waves, in particular into the necessity of friction.

Dressler~\cite{Dressler_roll-waves} went on to formalise these solutions, also providing continuous wave solutions to the corresponding viscous problem. He found that an entropically valid one-parameter family of roll-wave solutions exists with the specification of channel slope, resistance and wave speed. 
A solution is unique if also the wavelength is specified. 
This theory does however not explain why wave trains in nature are observed to consist of a relatively narrow band of wavelengths.

Richard and Gavrilyuk \cite{Richard_Gavrilyuk_roll_wave_model_comp_Brocks_exp} extended Dressler's roll-wave solutions
to account for turbulent shear and dissipation. 
Reynolds' stresses are related to enstrophy in the wave, 
providing wave-breaking as a model extension. 
Very good agreement was found with the experimental data of Brock's \cite{Brock_roll_wave_exp_PhD}, appropriately breaking off the sharp wave tip of the Dressler solutions.

Tougou and Tamada addressed 
the issue of roll-wave train stability in channel flow.
Both laminar \cite{Tamada_Tougou_wave_stability_laminar} and turbulent \cite{Tougou_wave_stability_turbulent} flows were considered. 
Their linear stability analysis indicated that  wavelengths observed in nature will be those of greatest linear stability.
Balmforth and Mandre~\cite{Balmforth_dynamics_of_roll_waves} also performed a stability analyses for shallow water roll-waves using a somewhat different technique. 

Miya et.\ al.\ \cite{Miya_roll_wave_gas_liq_channel} 
derived at profile solutions similar to those of Dressler for gas-liquid duct flows, also including shape factors for the velocity profiles.
Watson~\cite{Watson_wavy} in turn formulated such solutions for gas-liquid flow in pipes. 
This model had a form similar to that for channel flows, but with a geometrical complexity making it unsuited for analytical integration.
Algebraically explicit profile solutions are therefore unavailable for pipe flows.
The increased complexity of the equations also made solution uniqueness difficult to prove; this was instead assumed.

Comparisons between roll-wave experiments and predictions from finely resolved numerical representations have been made
by multiple authors. 
For instance, 
in the work of Brook et.\ al.\ \cite{Brook_roll_wave_Godunov_comp_dressler} the solutions of 
Dressler is compared to roll-wave simulation results of the shallow water equations using a second order Godunov method.
Holm{\aa}s \cite{Holmaas_roll_wave_model} compared a `pseudospectral' representation (using fast Fourier transformation) of the incompressible two-fluid pipe flow model with Johnson's experiments \cite{George_PhD}.  %the experiments of  Johnson's \cite{George_PhD,George_wave_model}. 
The Biberg model~\cite{Biberg_friction_duct_2007} for pre-integrated turbulent shear and velocity profiles was here incorporated with an extension to the interfacial turbulence closure to account for the effect of wave-breaking.
\\

A stability analysis similar to those given by Tougou and Tamada will here be 
applied to Watson's model for roll-waves in two-phase pipe flows.
Emphasis is placed on the stability of pressure propelled flows as opposed to gravity driven flows.
Stability predictions are compared to finely resolved numerical Roe scheme simulations.

\section{The Two-Fluid Model for Pipe Flow}
\label{sec:two_fluid_model}

\begin{figure}[h!ptb]%
\centering
\begin{minipage}[c]{.35\columnwidth}
\includegraphics[width=\columnwidth]{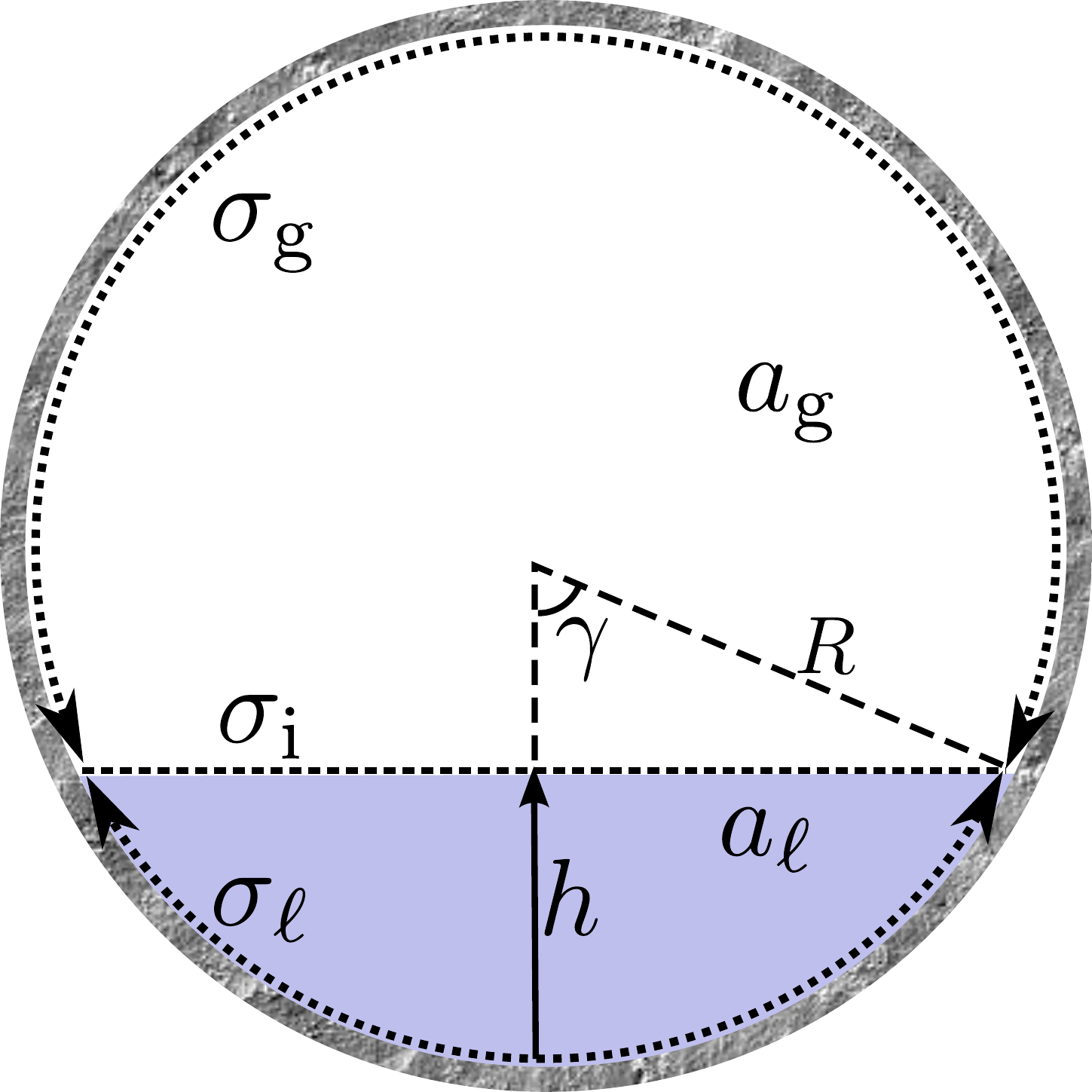}%
\end{minipage}
\hfill
\begin{minipage}[c]{.6\columnwidth}
\includegraphics[width=\columnwidth]{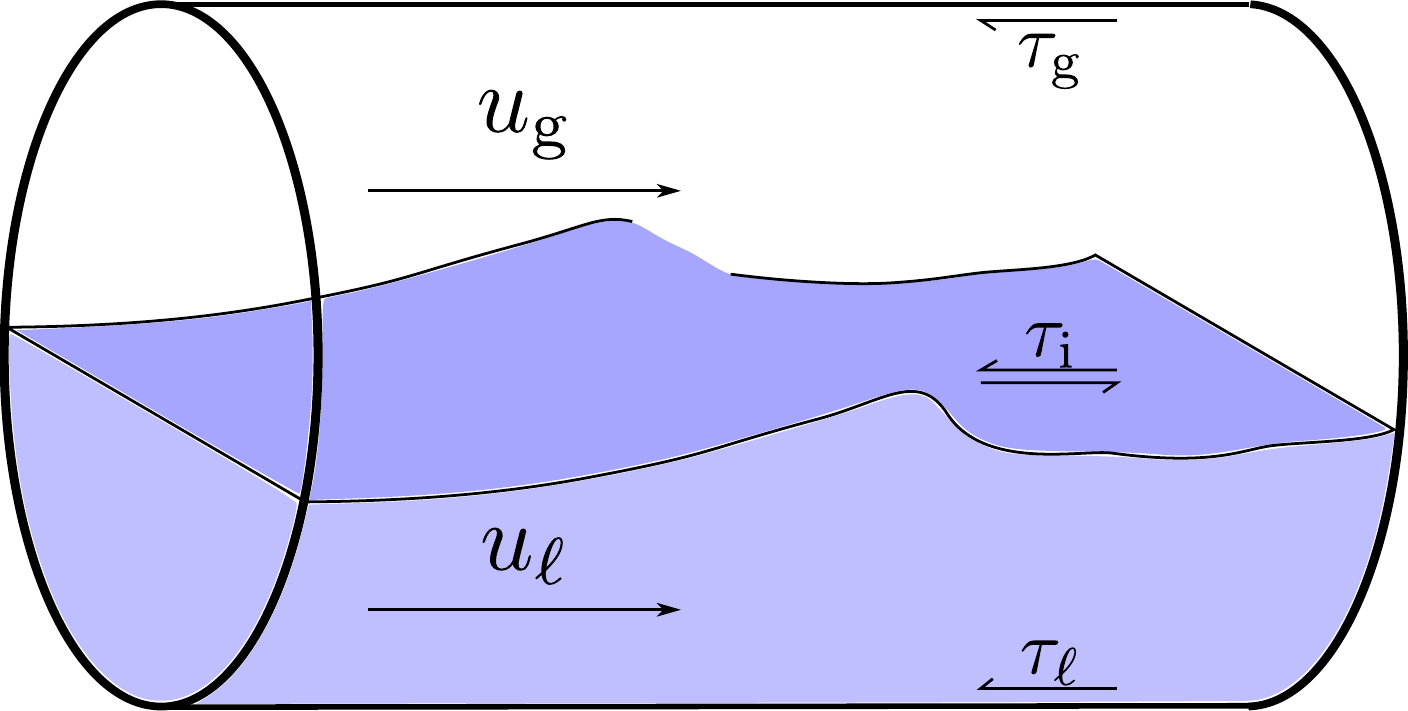}%
\end{minipage}
\caption{Pipe cross-section}%
\label{fig:cross_section}%
\end{figure}
Figure~\ref{fig:cross_section} illustrate the pipe geometry and some of the quantities appearing the two-fluid model.
The circular pipe geometry itself enters into the modelling through the relation between the level height $h$, the specific areas $\ak$ and the peripheral lengths $\sigma\k$ and $\sigma\_i$.
These are algebraically interchangeable:
\begin{align}
h &= \mc H(\al).
\label{eq:Alofh}
\end{align}
The inverse of the geometric function $\mc H$ can be explicitly expressed as
\begin{align*}
\al=\mc H\inv\of h &= R^2\br{\gamma - \nicefrac12 \sin 2\gamma},
&
\gamma &= \arccos\br{1-\frac h R}.
\end{align*}
$R$ is here the pipe inner radius and $\gamma$ the interface half-angle.
The perimeter lengths are 
\begin{align*}
\sigma\l &= 2 R \gamma,
&
\sigma\g &= 2R\br{\pi-\gamma},
&
\sigma\_i &= 2R\sin \gamma.
\end{align*}

The compressible, isothermal, four-equation two-fluid model for stratified pipe flow results from an averaging of the conservation equations across the cross sectional area. Field $\phaseindex$, occupied by either gas, $\phaseindex = \mr g$, or liquid, $\phaseindex = \ell$, is segregated from the other field. 
The model is commonly written
\begin{align*}
	&\br{\rho\k \wt a\k }_t+  \brac{\rho\k \wt a\k \wt u\k}_x = 0,\\
	& \br{\rho\k \wt  a\k \wt u\k}_t + \brac{\rho\k \wt a\k \wt u\k^2}_x + \wt a\k \wt p_{\mr i,x} + \rho\k g \cos\theta \, \wt a\k \wt h_x  = \wt s\k.
\end{align*}
Tildes have here been added to distinguish functions of the fixed frame coordinate $x$.
$ p\_i$ is the pressure at the interface, assumed the same for each phase as surface tension is neglected.\ $h$ is the height of the interface from the pipe floor, and the term in which it appears originates from approximating a hydrostatic wall-normal pressure distribution within each field. 
$\ak$ is the cross-section area within field $\phaseindex$, and $u\k$ and $\rho\k$ are the mean fluid velocity and density in these fields.
The momentum sources are 
$ s_k = \tau\k \sigma\k \pm \tau\_i  \sigma\_i/  - \ak\rho\k g \sin\theta$,
where $\tau$ is the skin frictions at the walls and interface, $\theta$ is the pipe inclination, positive above datum, and $g$ is the gravitational acceleration.

Both fluids flows are from here assumed incompressible. This allows for the system to be represented on conservative form.%
\footnote{It is sufficient for one of the fluids to be incompressible and the other to have a negligible level height term for this to be possible.}
%Reducing the momentum equations with their respective mass equations, dividing by the phase areas and finally eliminating the pressure term between phases yields
Reducing the momentum equations with their respective mass equations and eliminating the pressure term between them yields
\begin{equation}
\wt \bw_t + \wt\bff_x = \wt \bs,
\label{eq:base_model:abs}
\end{equation}
with components
\begin{align*}
\bw &= \big( \al, \diffk{\rho u}  \big)^T,
&
\bff &= ( \ql, \j )^T,
&
\bs & = (0,s)^T.
\end{align*}
Symbols for the flux and source components have here been defined and are
\begin{equation*}
\begin{gathered}
\qk = \ak u\k,
\qquad
\j =  \diffk{\rho  {u^2/2}} +  \my h,
\\
s = -\mx -\diffk{\tau \sigma/a} + \tau\_i \sigma\_i \br{1/\al+1/\ag}.% \br{\al\inv+\ag\inv}%\sumk \ak\inv
\end{gathered}
\end{equation*}
The shorthand
\[
	\diffk \cdot = (\cdot)\l-(\cdot)\g
\]
is useful throughout.
Constant weight parameters have been grouped into
$\mx = \diffk\rho g \sin \theta$ and $\my = \diffk\rho g \cos \theta$.
Although really derivatives therefrom, these equations are here simply referred to as base mass and momentum equations.

The identities
\begin{align}
\al+\ag &= \area, 
& 
\ql+\qg &= \Qm,
\label{eq:sum_a_au}
\end{align}
where the latter conditions has been obtained from summing the mass equation over the two phases, finally close the base model.
Both $\area$ and $\Qm$ are parametric -- the former may be made to vary in space according to the geometry and the latter in time according to the mixture rate imposed upon the system.

\section{The Steady Roll-Wave Solutions}
\label{sec:steady_rollwave}

The profile equation is obtained by searching for system solutions which are steady within a moving reference frame. 
A chance in spatial coordinates $x \rightarrow \x = x - C t$ is performed, $C$ being a constant translation velocity equal to the wave celerity. 
The base model \eqref{eq:base_model:abs} with the new relative variables $\psi\of{\x,t}=\tilde \psi\of{x,t}$ then reads
\begin{equation}
\bw_t + \bff\_{r,\x} = \bs
\label{eq:base_model:rel}
\end{equation}%
The coordinate transformation brings about a new term which has been absorbed into the convection term. 
Its net effect is to make convection flux velocities relative to $\x$, \ie
\[
\ukr = u\k-C.
\]
$\bff\_r = (\qlr, \jr)^T$
 are the fluxes with $\ukr$ replacing $u\k$.

\subsection*{Profiles}
Denoting the properties of steady profile solutions with upper-case symbols, the profile equation is obtained by substituting $\psi\of{\x,t}\rightarrow\Psi\of{\x}$ in \eqref{eq:base_model:rel}. 
As the transient term drops out 
 the mass equation component and \eqref{eq:sum_a_au} reviles
\begin{equation}
\Qkr = \Ak \Ukr=\const..
\label{eq:Qr}
\end{equation}
Regarding $\Qr$, $\Jr$ and $\S$ as functions of $\Al$ subjected to \eqref{eq:Qr}, the chain rule applied to the momentum equation component of \eqref{eq:base_model:rel} yields the profile equation
\begin{equation}
\dAl =  \frac {\S\of\Al}{\N\of \Al}.
\label{eq:profile}
\end{equation}
The numerator is
\[
	\N \equiv \ddiff\Jr\Al= \Drhogcos 
		- \br{\rho\, \Ur^2}\m
\]
 with $\mc H' = \ddiff {\mc H} \Al$.
Another useful operator
\[
\psi\m = \frac{\psi\l}\Al+\frac{\psi\g}\Ag
\]
has here been introduced.
\\

A transition from subcritical flow, ($\sign (\lambda^-) \neq \sign(\lambda^+)$) to supercritical flow ($\sign (\lambda^-) = \sign(\lambda^+)$)
is necessary for the formation of periodic hydraulic jumps. 
In fact, the root of the slow characteristic $\lambda^-$ (see \eqref{eq:eigenvalues}) coincides with the root of $\N$ in what is known as the `critical point' $\Alnil$. 
For a steady state to be possible, and \eqref{eq:profile} to be integrable, $C$ must be such that also $\S$ has a root at the critical point $\Alnil$, that is $C: \; \N\of{\Alnil;C} = \S\of{\Alnil;C} = 0$.
Supercritical flow is found in the range $\Al < \Alnil$ and subcritical in the range $\Al > \Alnil$.
Profile solutions are obtained by numerically integrating \eqref{eq:profile} inversed, \ie, 
$\xi =\int\! \N/\S \,\mr d\Al$.

\subsection*{Shocks}
\begin{figure}[h!ptb]%
\centering
\includegraphics[width=\imscale\columnwidth]{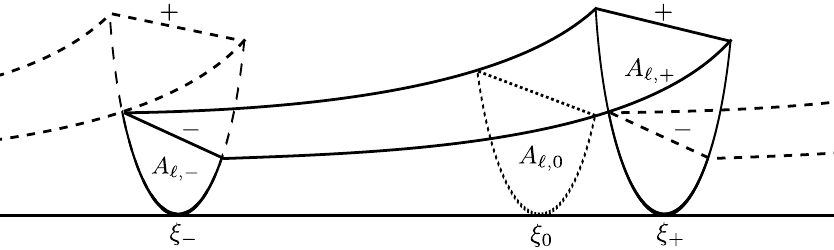}%
\caption{Illustration of the steady wave solution.}%
\label{fig:steady_roll_wave}%
\end{figure}

Entropically valid solutions of \eqref{eq:profile} are monotone, 
but periodic wave trains are possible by piecewise connecting profile solutions through shocks.
Integrating the conservation equations \eqref{eq:base_model:rel} thinly over a shock front reviles that $\qlr$ and $\jr$ are shock invariants, \ie\ 
\begin{equation}
	\diffpm{\bff\_r}=\bm 0,
	\label{eq:shock_condition}
\end{equation}
where the `$+$' and `$-$' respectively constitutes the right and left limits of a shock front. 
$\psi_\pm = \psi\of{\x_\pm}$ and $\diffpm \cdot= (\cdot)_+ - (\cdot)_-$.
The first shock condition component is trivially achieved in the steady wave train by virtue of \eqref{eq:Qr}, 
while the momentum flux shock condition reads
\begin{equation}
	\diffpm{\tfrac 12\! \diffk{\rho \,\Ur^2} + \my \,H } = 0.
\label{eq:shock_condition:momentum}
\end{equation}
Wave profiles are assumed identical along the wave train.
The wave profile is therefore repeating such that for any integer $n$ we have $\Psi\of{\x+n\lambda}=\Psi\of\x$, $\lambda$ being the wavelength
(not to be confused with the characteristics $\lambda^\pm.$)
The left and right limit states are therefore the states at the tip and tail of the same profile solution.
Condition~\eqref{eq:shock_condition:momentum} links these states.
Figure~\ref{fig:steady_roll_wave} shows a schematic.

For a set of fluid and geometry parameters, equations \eqref{eq:sum_a_au}, \eqref{eq:profile} and \eqref{eq:shock_condition} generate a one-parameter family of wave train solutions.

\section{The Linear Stability of Roll-Wave Trains}
The procedure adopted for analysing the linear stability of the steady roll-wave trains is inspired by Tougou and Tamada~\cite{Tougou_wave_stability_turbulent, Tamada_Tougou_wave_stability_laminar}, and Balmforth and Mandre \cite{Balmforth_dynamics_of_roll_waves}.

\subsection*{The Disturbace Function}

Perturbations are imposed on the steady state:
\begin{equation}
\psi\of{\x,t} = \Psi\of\x + \wh \psi e^{\omega t},
\qquad \psi\in\{\a\k,u\k,\eps\}
\label{eq:linarizing}
\end{equation}
where the \textit{pulsation} $\omega$ is a (complex) constant.
\begin{subequations}
A disturbance function $\f\of{ \x}$ is then introduced and defined by
\begin{align}%
\ap\l &=  \df.%\of \x ,
\label{eq:hat:al}
\intertext{%
Inserting these definitions into the mass equation component of \eqref{eq:base_model:rel} and integrating once yields
}
\up\l &= -\frac{\omega \f + \df\,\Ulr}{\Al}.
\label{eq:hat:ul}
\intertext{%
An arbitrary integration constant has here been dropped.
Further, due to the identities 
\eqref{eq:sum_a_au},
the following holds:
}
\ap\g &= -\df, %-\ap\l,
\label{eq:hat:ag}
\\
\up\g &= \frac{\omega \f + \df\,\Ugr}{\Ag}.
\label{eq:hat:ug}
\end{align}%
\label{eq:hat}%
\end{subequations}%
The disturbance function $\f$ is here that eigenfunction which will provide a constant $\omega$ in \eqref{eq:linarizing} throughout the wave.

The friction closure is kept unspecified. Following the common practice in stability studies of pipe flows \cite{Barnea_Structural_instability_combined_VKH_nonlinear},
$\S$ is expressed with the discharges (analogous to superficial velocities) as separate variables, \ie, $\S\of\Al = \mcS\of{\Al,\Q\l,\Q\g}$. %, and linearised. 
Subject to \eqref{eq:Qr}, the chain rule yields
\begin{equation}
\S' = 
\ddiff \S \Al =
\diffA + C \br \diffU,
\label{eq:dSdAl_tot}
\end{equation}
where
\begin{align*}
\mcS_{\Al} = &\br{\pdiff \mcS\Al}_{\!\!\Q\l,\Qg}\!\!\!\!\!\!\!\!\!\!\!\!\!, &
\mcS_{\Ql} = &\br{\pdiff \mcS\Ql}_{\!\!\Al, \Qg}\!\!\!\!\!\!\!\!\!\!\!\!\!, &
\mcS_{\Qg} = &\br{\pdiff \mcS\Qg}_{\!\!\Al, \Ql}\!\!\!\!\!\!\!\!\!\!\!\!\!.
\end{align*}

Substituting $\wh\psi$ from \eqref{eq:hat} into the momentum equation component of \eqref{eq:base_model:rel} and linearising results in
\begin{equation}
\cftwo \ddf  +\cfone \df + \omega  \cfnil\f = 0,
\label{eq:disp_f}
\end{equation}
where
\begin{align*}
\cftwo &= \N,
\\
\cfone  &=  3\, \br{ \rho\, \Ur^2 \frac{\dA}{\A}}\m - 2\,\omega \br{\rho\, \Ur }\m
		-\S'
\\
\cfnil &=  2\,\br{ \rho\,\Ur \frac{\dA}{\A}}\m - \omega\rho\m  +\diffU.
\end{align*}
$\N$, being the denominator of the profile equation \eqref{eq:profile}, equals zero at the critical point $\Alnull$, which is a double root. It then follows from \eqref{eq:disp_f} that
\begin{equation}
\dfnil = - \omega \f_0  \left.\frac{\cfnil}{\cfone}\right|_0.
\label{eq:dfdAl0}
\end{equation}
L'H{\texthat o}pital's  rule 
$
\A_{\ell,\x,0} = \N'\of{\Alnil}/\S'\of{\Alnil}
$
may here be used for evaluating $\dfnil$.
\\

Integration of \eqref{eq:disp_f} can be done with either $\Al$ or $H$ as independent variable. 
Because only the inverse $\Al = \mc H\inv\of H$ is an explicit function, $H$ is used  in the presented numerical experiments.
The chain rule yields
\begin{align*}
\df &= \dHdAl \dAl \ddiff \f H,
\\
\ddf &= (\dHdAl \dAl)^2 \dddiff \f H + \br{\ddHdAl \dAl^2 + \dHdAl \ddAl} \ddiff \f H.
\end{align*}

The double derivative $\ddAl$ has appeared above and is
\begin{equation*}
\ddAl = \brac{ \frac {\S'}\S - \frac {\N'}\N }\dAl^2 
\end{equation*}
with
\begin{equation*}
\N' = 
\my \,\ddHdAl
+3\diffk{\rho\br{\Ur/\A}^2 }
\end{equation*}
and $\S'$ from \eqref{eq:dSdAl_tot}.
Derivatives of the the geometric function $\mc H$ are
\begin{align*}
\dHdAl &= 1/\sigma\_i,
&
\ddHdAl &= 4\br{h-R} {\dHdAl}^4.
\end{align*}

\subsection*{Shocks}
A shock, unperturbed travelling with the celerity $C$, will be displaced by a length $\eps\of t = \epsp \,e^{\omega t}$.
The perturbed shock speed is then $c = C+\omega \eps$. %$c = C+\deps$.
Shock conditions \eqref{eq:shock_condition} must now be evaluated at the location $\xshock+\eps$ of the perturbed shock, $\xshock$ being an unperturbed shock position. 
The left and right shock states are expressed through Taylor-expansions, $\Psi\of{\xshock+\eps} = \Psi\of{\xshock}+\eps\Psi_\x\of{\xshock}$, disregarding all higher-order terms. % in $\wh \psi$. 
We get
\begin{subequations}
\begin{gather}
\diffpm{ \Al\ulpr + \ap\l\Ulr } = 0,
\label{eq:shock_hat:mass}
\\
\diffpm{ \diffk{\rho \,\Ur\br{ \epsp\, \dU +  \upr  }} +\my {\dHdAl}\br{\epsp \dAl + \ap\l } }=0,
\label{eq:shock_hat:mom}
\end{gather}
\label{eq:shock_hat}%
\end{subequations}%
with $\ukpr = \up\k-\omega\epsp$.
Inserting the disturbances \eqref{eq:hat} into \eqref{eq:shock_hat} and eliminating $\epsp$ then  %(defined in \eqref{eq:linarizing}) 
yields a perturbed shock condition
\begin{equation}%
\diffpm{\N \df - \omega \br{\rho\,\Ur}\m \!\f }
- \frac{\bdiffpm \f}{\diffpm \Al} \diffpm{\S - \omega \diffk{\rho\,\Ur} }
= 0
\label{eq:shock_hat:mom_LR}%
\end{equation}%
linking the eigenfunction value $\f_-$ at the right-wave tail to the value $\f_+$ at the left-wave front.

\subsection*{Stability of Wave Trains}

Solutions of the eigenfunction problem \eqref{eq:disp_f}--\eqref{eq:dfdAl0} may be written as the family
\begin{equation}
\f = \b \F, \qquad \F\of{\x_0} = 1,
\label{eq:F_def0}
\end{equation}%
$\b$ being an arbitrary constant. 

Let $\n$ be the sequential count of roll-waves down along a wave train.
The origin of $\n$ is irrelevant.
$\F$ is also  $\lambda$-periodic, $\F\of{\x + n\lambda}=\F\of\x$, because $\F$ depends only on the steady wave solution.
Indexing  individual values of $\b$ according to the wave count, one may express the disturbance along the entire wave train as 
\begin{equation*}
\f\of{\x+\n \lambda}=\b_\n \F\of{\x}; \qquad 0 \leq \x \leq \lambda, \quad\n\in \mathbb Z.
\end{equation*}%
Consider a shock connecting wave $\n$ to wave $\n+1$. The left shock state may be written $\f_+ = \b_\n\F\of{\x_+}$ and the right $\f_- = \b_{\n+1}\F\of{\x_-}$.
%Equation~\eqref{eq:shock_hat:mom_LR} now relates the only wave-specific variables, $\b_\n$ and $\b_{\n+1}$, by an amplification only dependent of the unperturbed wave solution. % and $\F$. 
The amplification across each shock is therefore repeating in both $\n>0$ and $\n<0$ directions along the wave train; 
for a disturbance to be bounded in a infinite or periodic spatial domain the amplification must have the form
$
\b_{\n+1} =\b_\n e^{i \,2\pi/\Gm} ,
$
$\Gm$ being the number of roll-waves in a disturbance period.
A periodic disturbance then has the form
\begin{equation*}
\f\of{\x+\n\lambda} = \b_0\F\of\x e^{i \, 2\pi  \frac \n\Gm}; \quad 0\leq\x\leq\lambda, \quad \n\in\mathbb Z, \quad \Gm \in \mathbb N.
\end{equation*}

The stability of a wave train is determined by searching for $\omega$-roots of \eqref{eq:shock_hat:mom_LR} for given values of $\Gm$. % with $\G\in[0,\pi]$. 
Roots in the right half of the complex plane, $\Re(\omega)>0$, are unstable.
\\

$\F$ from \eqref{eq:disp_f}, \eqref{eq:dfdAl0} and \eqref{eq:F_def0} has solutions obtainable with a Frobenius method
about the critical point.
However, the geometric relations and any extensive friction closure entail Taylor expansions which make such formulations impracticable. 
$\F$ is instead determined by integrating  \eqref{eq:disp_f} numerically, using a Rounge-Kutta ODE solver. Integration is performed from
$\big(H_0+\delta_H,1+\delta_H \ddiff \F H\big|_0\big)$ to $\big(H_+, \F_+\big)$ 
on the subcritical side, and from
$\big(H_0-\delta_H,1-\delta_H \ddiff \F H\big|_0\big)$ to $\big(H_-, \F_-\big)$ 
on the supercritical side, $\delta_H$ being a tiny height step for the purpose of avoiding numerical $0/0$-issues.

Partial derivatives of friction closures of arbitrary complexity may be computed in a discrete manner
\[
\mcS_{\Al} = \big(\mcS\of{\Al+\h\delta_\A,\ql,\qg}-\mcS\of{\Al-\h\delta_\A,\ql,\qg}\big)/\delta_\A,\quad \text{etc..}
\]

The stability analysis for open channel roll-waves, such as presented by cited authors, is regained by choosing 
$\mc H\of \al = \al/d$,
$\area = d^2$,
$\rho\g =\tau\g = \tau\_i = 0$.
%The value assigned to $\rho\l$ is then arbitrary.

\section{The Stability of Uniform Stratified Flow}
The well-known `viscous Kelvin-Helmholtz' (VKH) criterion \cite{Barnea_Structural_instability_combined_VKH_nonlinear} is often used for predicting  whether or not the flow regime is uniformly stratified. 
This analysis proceeds by inserting a Fourier disturbance mode in the uniform flow solution $\S = 0$. 
In absence of surface tension, the resulting dispersion equation can, despite the complicated expressions usually presented, be written
\begin{equation}
\N +\frac{i}{k} \S' = 0,
\label{eq:VKH_dispersion_eq}
\end{equation}
$k = 2\pi/\lambda$ being the wavenumber.
The celerity $C$ appearing in $\N$ and $\S'$ is here a complex value, related to the wave growth in the same manner at the pulsation in \eqref{eq:linarizing} by $\omega = -ik C $.

Notice that the condition for marginally stable flow, \ie, when $C$ is real, will simply read
$
\S = \S' = \Jr' = 0,
$
with $\Jr' < 0$ resulting in wave growth.
Note further that this can be inferred directly from \eqref{eq:profile} as a zero-amplitude roll-wave. 
All points are then critical points ($\N=\S=0$) where the profile slope is at an equilibrium with respect to $\Al$ ($\S'=0$).

\section{Numerical Experiments}
\label{sec:exp}
Stability predictions are in this section compared with direct simulations of the base model \eqref{eq:base_model:abs}. A Roe scheme, presented in \ref{sec:Roe}, is implemented for the purpose. 
Simulations are carried out with periodic boundary conditions. This means that the globally average area fractions stay constant in time.  $\Qm$ is also kept constant. %$\an{\wt \q\k}$ will change until a steady state is obtained.
Direct simulations are carried out over domains large large enough to avoid the boundaries influencing the statistics. 
A small pointwise random disturbance is issued to the uniform steady initial conditions.

\begin{figure}[h!ptb]%
\centering
\begin{subfigure}{.51\columnwidth}
\includegraphics[width=\columnwidth]{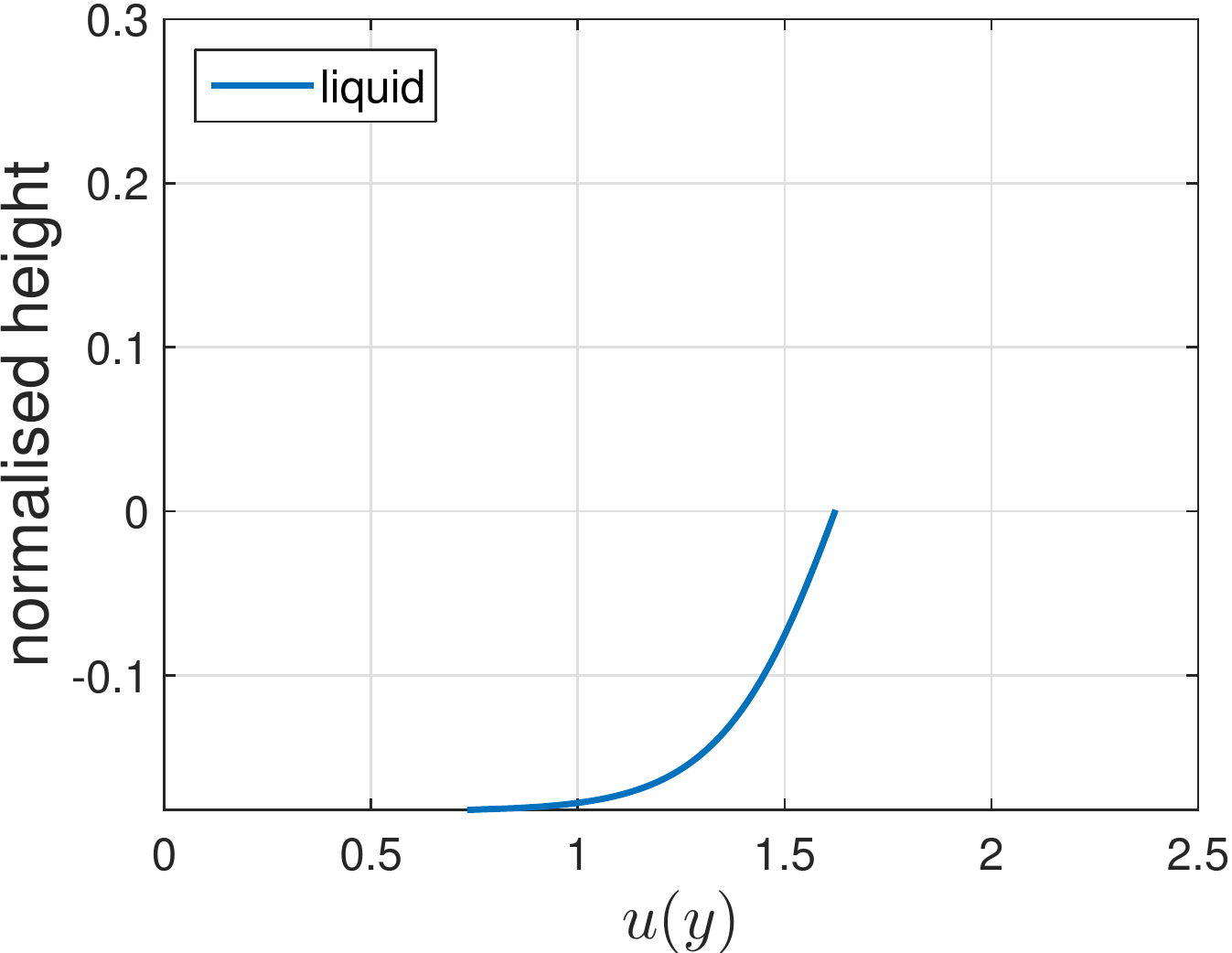}%
\caption{Free surface flow}
\label{fig:Biberg_profile:free_surface}%
\end{subfigure}
\hfill
\begin{subfigure}{.47\columnwidth}
\includegraphics[width=\columnwidth]{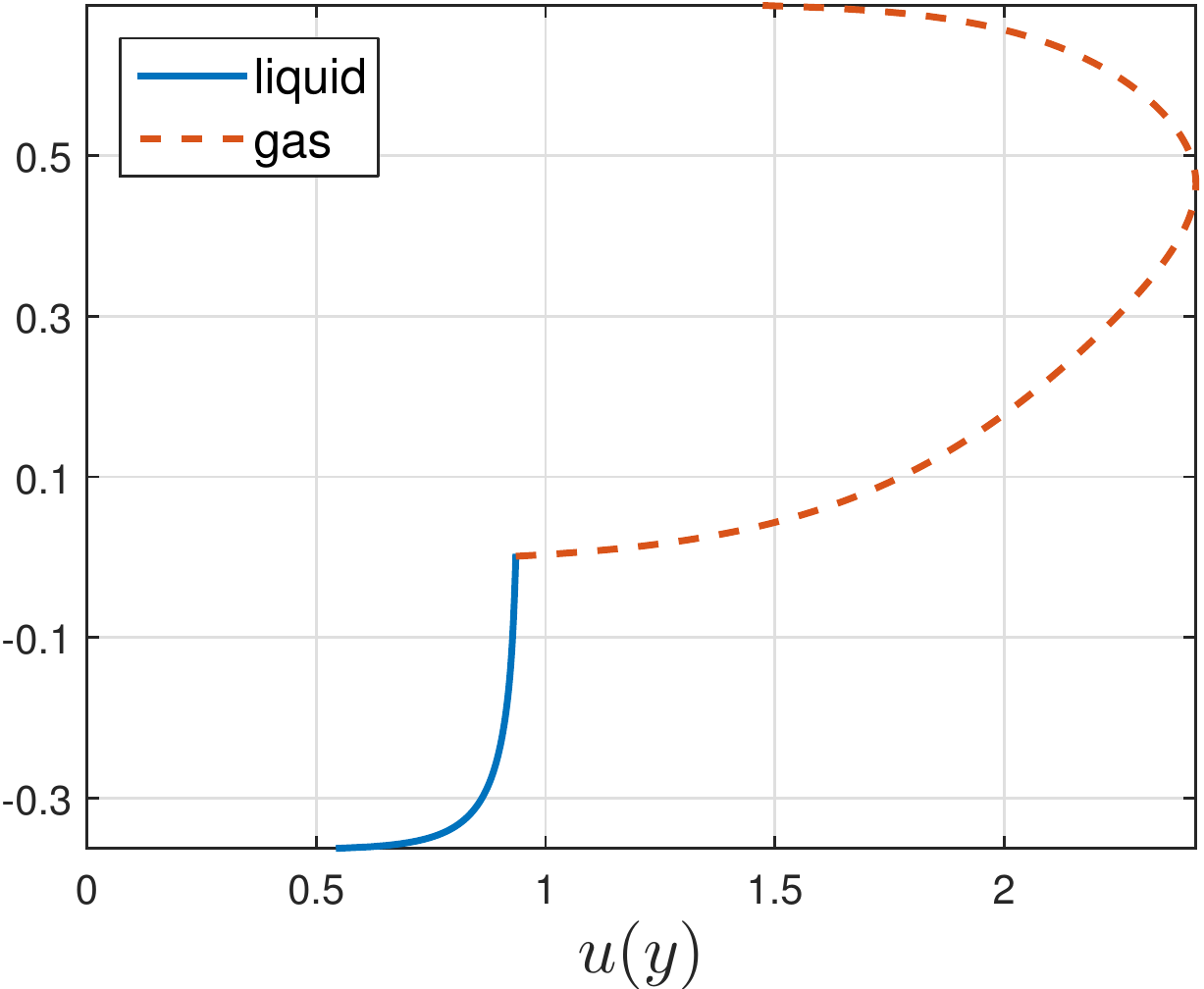}%
\caption{Gas-liquid flow}
\label{fig:Biberg_profile:two_phase}%
\end{subfigure}
\caption{Example velocity profile pre-integrated in the Biberg friction model.  }%
\label{fig:Biberg_profile}%
\end{figure}

The friction closures $\tau\k$ and $\tau\_i$ in the numerical tests are provided by the Biberg friction model as presented in \cite{Biberg_friction_duct_2007}. 
A quick summary of this model now follows:
Classical turbulent  boundary layer principles are used to model the gas and liquid velocity profiles in a duct cross section. 
The interface is modelled as a moving boundary with an initial turbulence level, representing smaller interface waves, etc..
Figure~\ref{fig:Biberg_profile} shows examples of such velocity profiles for free surface and two-phase flows. 
These profiles are integrated up to yield algebraic expressions that couple wall and interfacial frictions to the average phase velocities and the interface heights. 
Friction correlations for duct flow are then correlated to the well-known Colebrook-White formula for single-phase pipe flow, which is in turn used to extend the closure 
into formulae for the pipe geometry.
\\

\begin{table}[h!ptb]
\centering
\begin{tabular}{|rl|rl|}
\hline
liquid density&				$\rho\l$	&998&			$\unitfrac[]{kg}{m^3}$\\
gas density&				$\rho\g$	& 50&			$\unitfrac[]{kg}{m^3}$\\
liquid dynamic viscosity&	$\mu\l$ 	& 1.00\Em3&		$\unit{Pa\:s}$\\ %\unitfrac[]{kg}{ms}
gas dynamic viscosity&		$\mu\g$ 	& 1.61\Em5&	$\unit{Pa\:s}$ \\
internal pipe diameter&		$d$ 		& 0.1&			$\unit[]m$	\\
wall roughness&							&2\Em5& 		$\unit[]m$ \\\hline
\end{tabular}
\caption{Fixed parameters.}
\label{tab:parameters}
\end{table}

\begin{figure}[h!ptb] %
\centering
\begin{subfigure}{\imscale\columnwidth}%
\caption{$\Omega$ stability map. Contour lines show $\Omega$ values forming roots in \eqref{eq:shock_hat:mom_LR} at specified wavelengths. Normalised wavelengths $\lambda/d$ are numerically indicated within the respective contour lines.
}%
\includegraphics[width=\columnwidth]{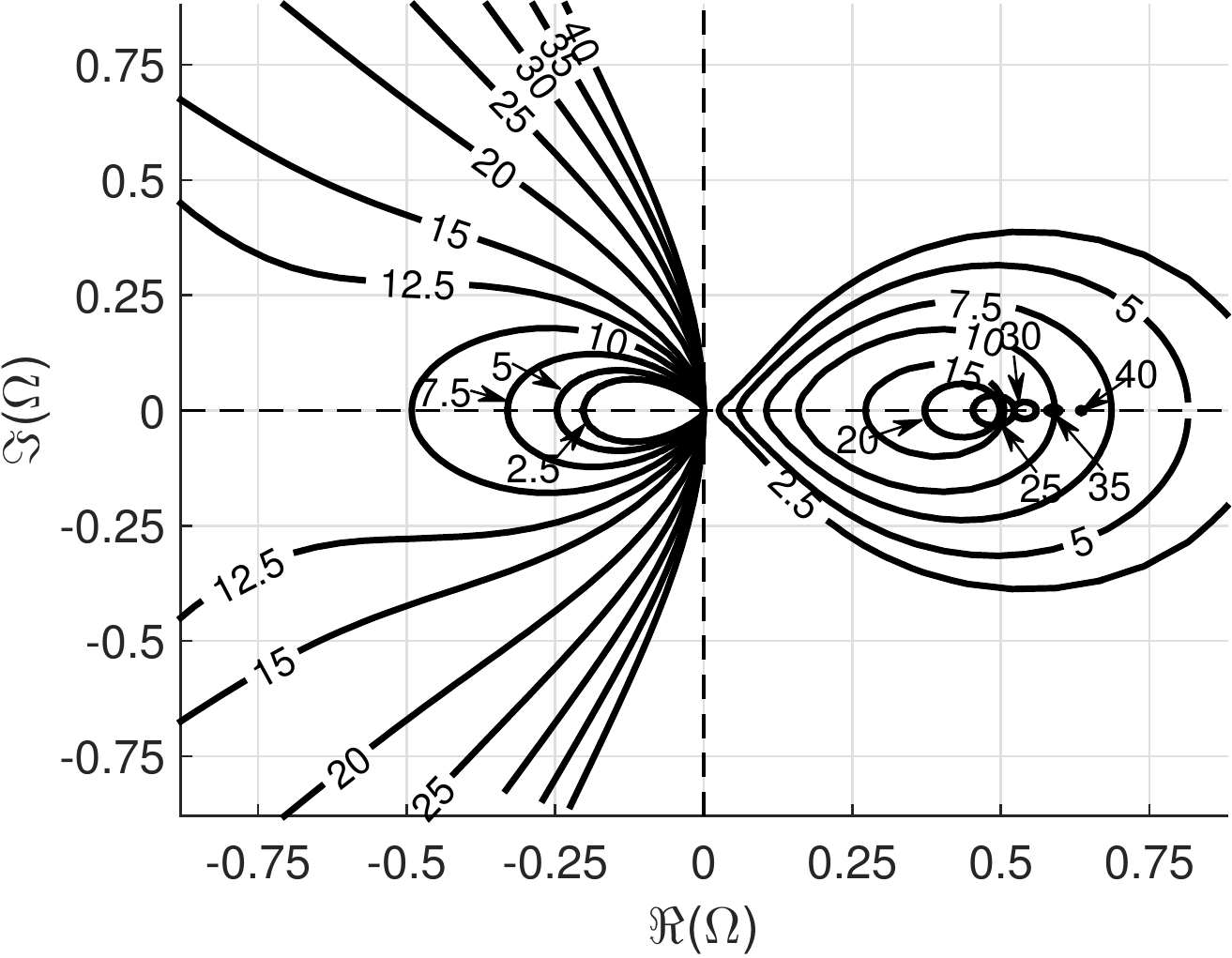}%
\label{fig:1deg:wavelength:stab}%
\end{subfigure}%
\\
\begin{subfigure}{\imscale\columnwidth}%
\caption{Direct simulation.  $10\,000$ grid cells in a $2\,000\,d$ long simulation domain.}%
\includegraphics[width=\columnwidth]{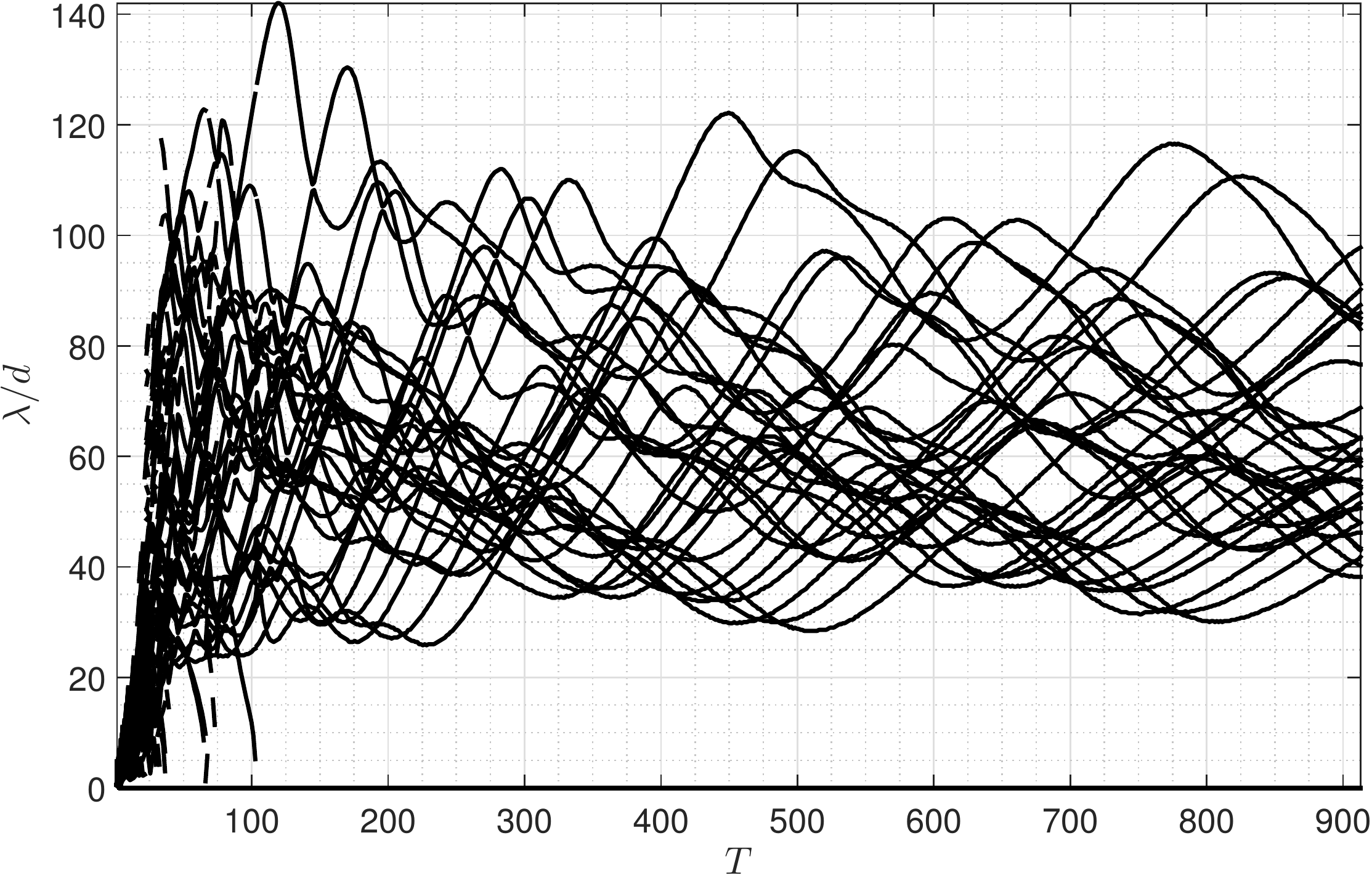}%
\label{fig:1deg:wavelength:Roe}%
\end{subfigure}%
\caption{$\theta = 1\degree$, $\USL = \unitfrac[0.125]ms$, $\USG = \unitfrac[3.50]ms$}%
\label{fig:1deg:wavelength}
\end{figure}

\begin{figure}[h!ptb]%
\centering
\begin{subfigure}{\imscale\columnwidth}%
\caption{$\Omega$ stability map, cf. caption of Figure~\ref{fig:1deg:wavelength:stab}
}%
\includegraphics[width=\columnwidth]{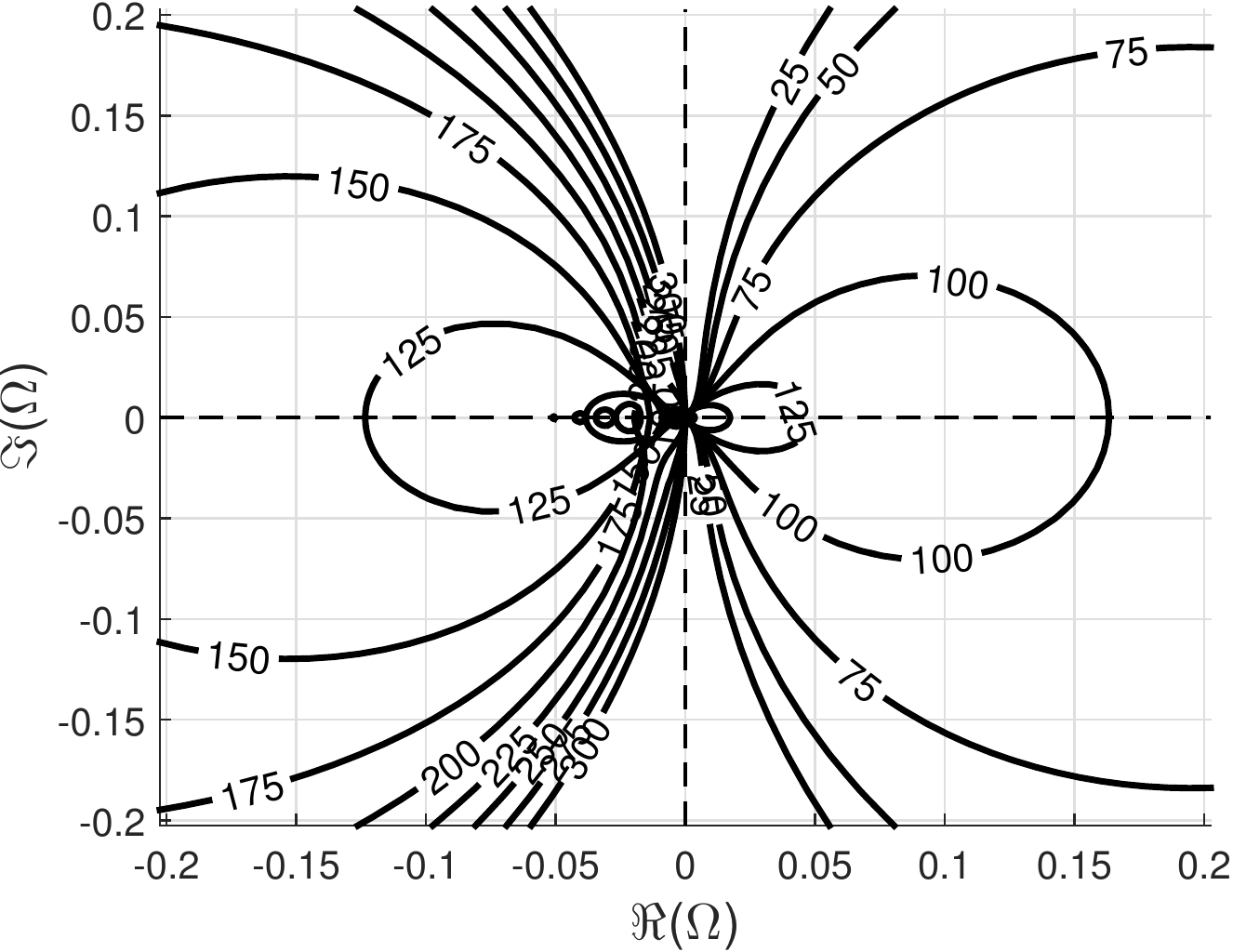}\\%
\includegraphics[width=\columnwidth]{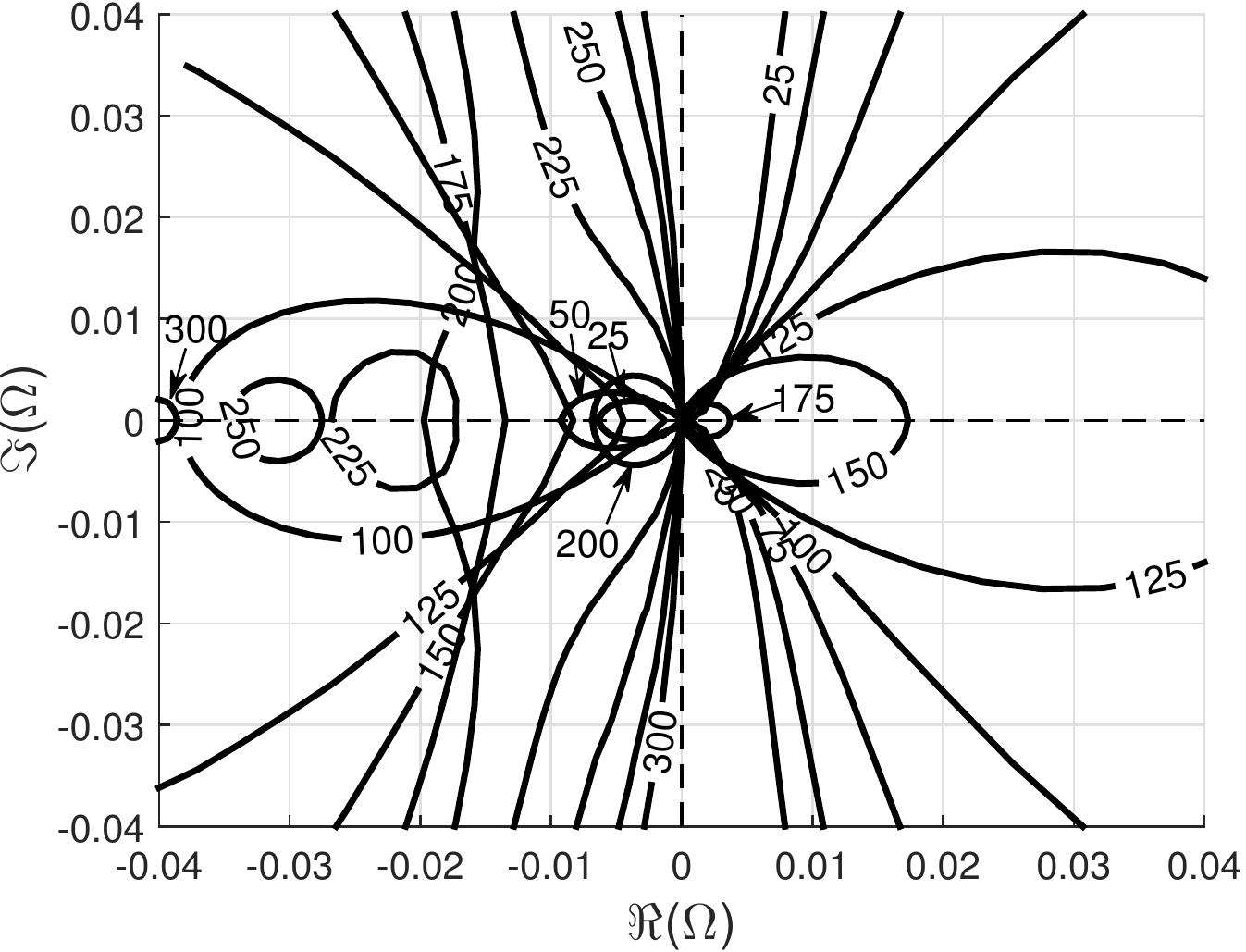}%
\label{fig:0deg:wavelength:stab}%
\end{subfigure}
\\
\begin{subfigure}{\imscale\columnwidth}%
\caption{Direct simulation. $10\,000$ grid cells in a $10\,000\,d$ long simulation domain.}%
\includegraphics[width=\columnwidth]{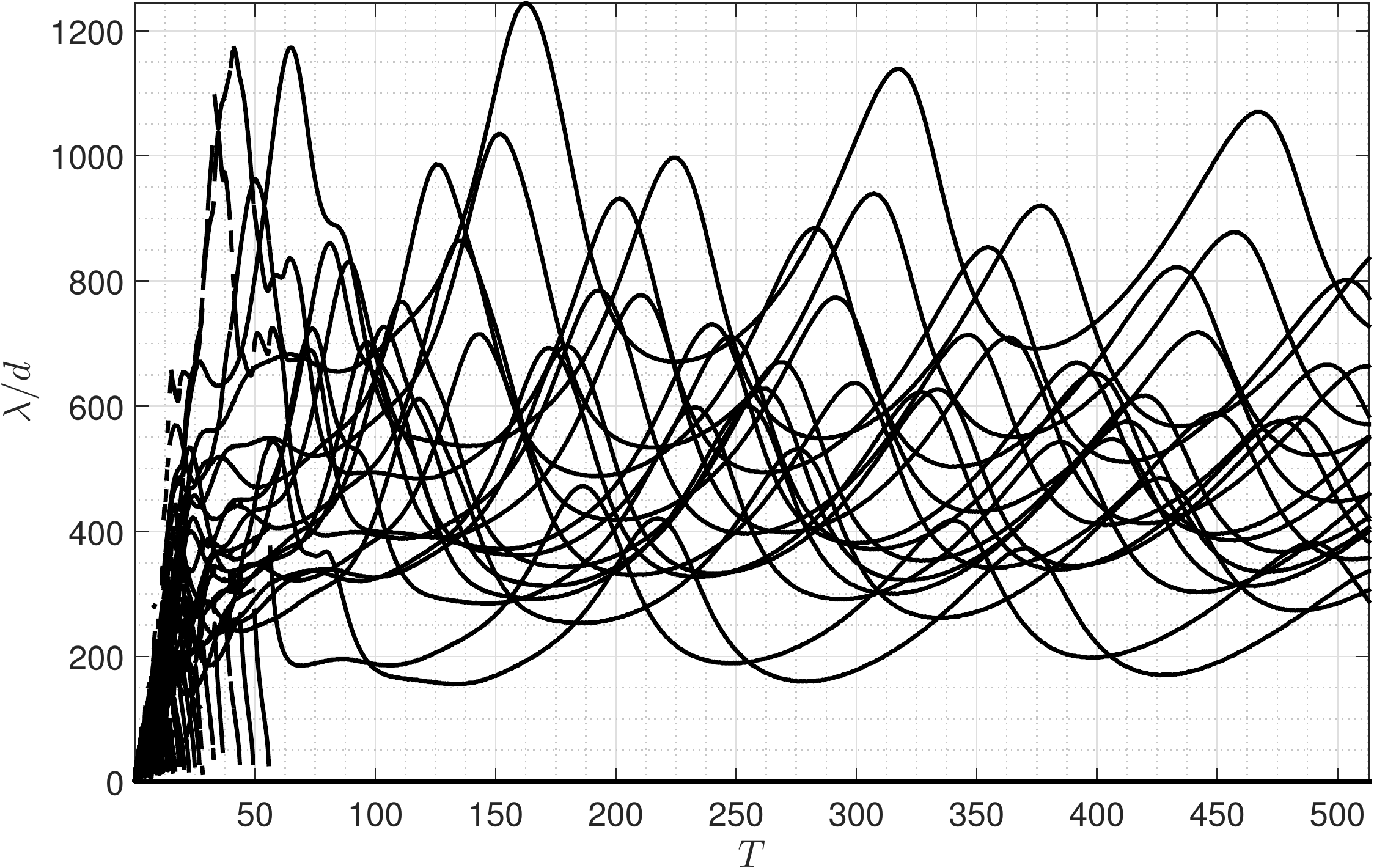}%
\label{fig:0deg:wavelength:Roe}%
\end{subfigure}%
\caption{$\theta = 0\degree$, $\USL = \unitfrac[0.35]ms$, $\USG = \unitfrac[1.00]ms$}%
\label{fig:0deg:wavelength}%
\end{figure}

\begin{figure}[h!ptb]%
\centering
\begin{subfigure}{\imscale\columnwidth}%
\caption{$\Omega$ stability map, cf. caption of Figure~\ref{fig:1deg:wavelength:stab} }%
\includegraphics[width=\columnwidth]{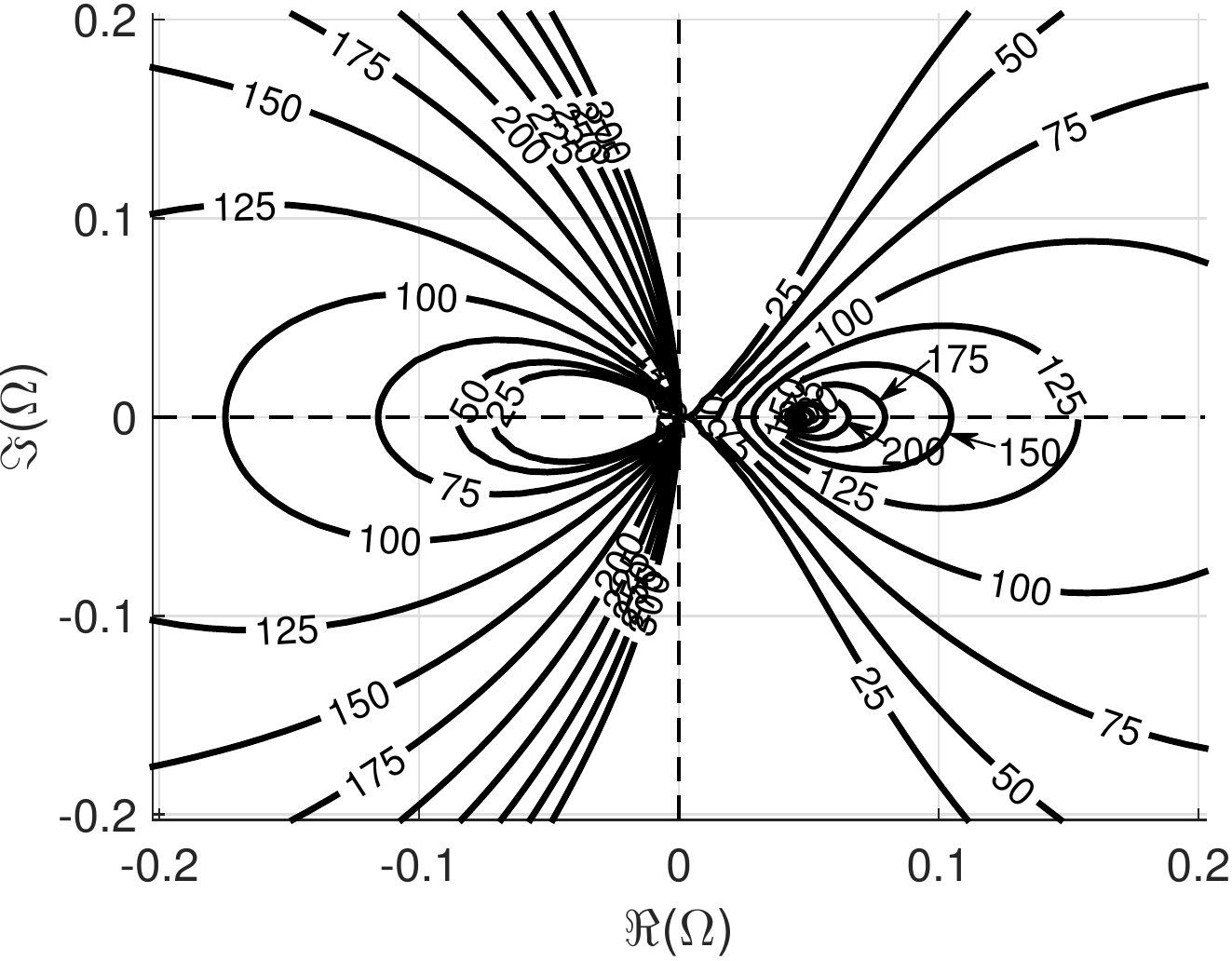}%
\label{fig:0deg0rhog:wavelength:stab}%
\end{subfigure}%
\\
\begin{subfigure}{\imscale\columnwidth}%
\caption{Direct simulation.  $10\,000$ grid cells in a $10\,000\,d$ long simulation domain.}%
\includegraphics[width=\columnwidth]{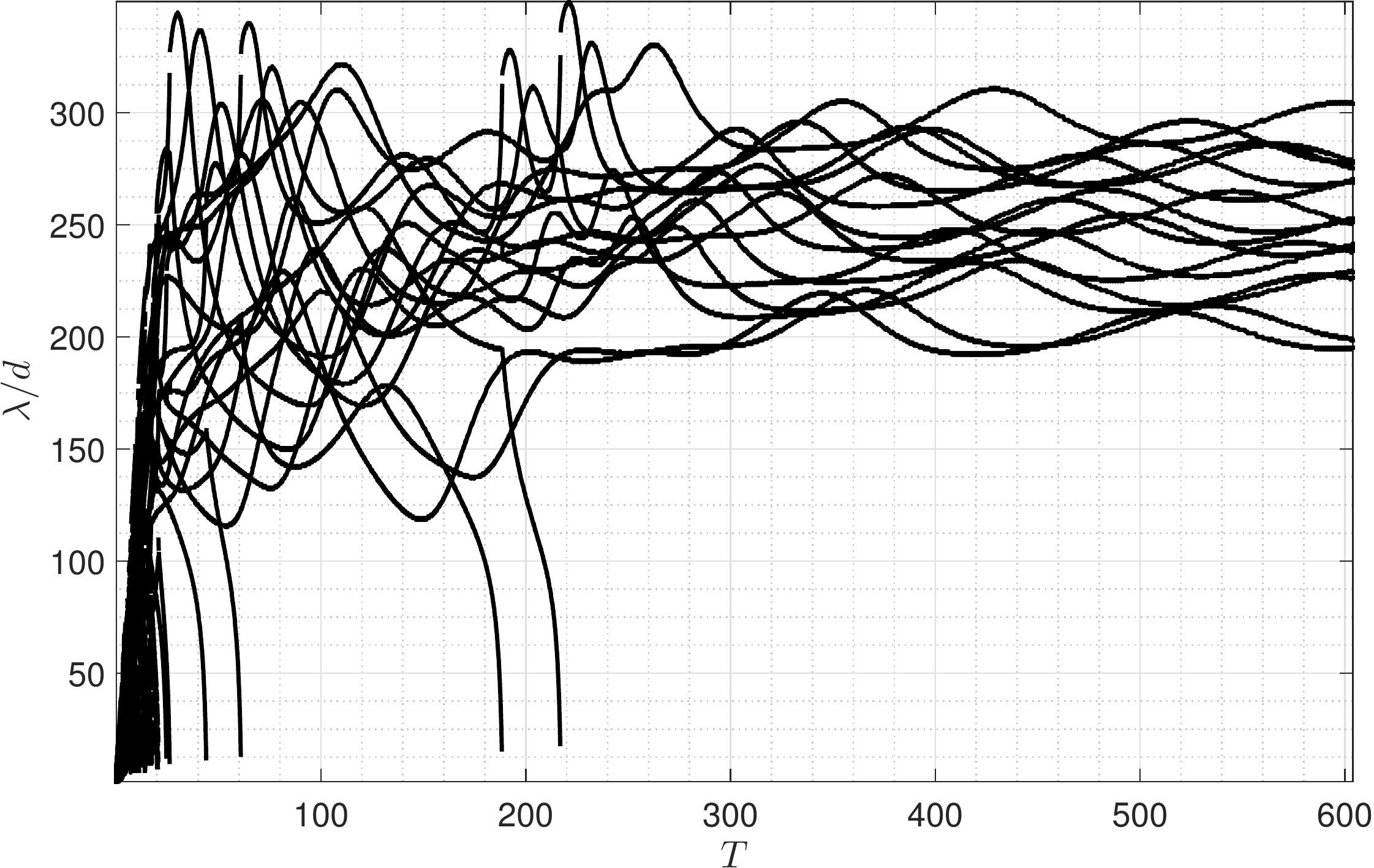}%
\label{fig:0deg0rhog:wavelength:Roe}%
\end{subfigure}%
\caption{Free surface flow; $\theta = -1.27\degree$, $\USL = 0.35\unitfrac ms$. VKH growth rate $\Re(\omega\_{VKH})$ is the same as in Figure~\ref{fig:0deg:wavelength}.}%
\label{fig:0deg0rhog:wavelength}
\end{figure}

Figures~{\ref{fig:1deg:wavelength} -- \ref{fig:0deg0rhog:wavelength}} present wavelength predictions from the stability analyses and direct simulations. % for two cases. 
Fixed parameters are given in Table~\ref{tab:parameters} and correspond to those used in \cite{Holmaas_roll_wave_model} and \cite{George_PhD}.
Flow rates are chosen 
low enough for the waves to grow from a random disturbance without ever breaching the cross section or flow characteristics turning complex.%
\footnote{
This range of flow conditions is rather narrow, with a weak growth rate.
Holm{\aa}s, in the cited paper, managed to simulate a much wider flow range within the roll-wave regime by expanding Biberg's friction closure with a wave breaking model. 
This model does however introduce level height derivative terms into the source model, altering the discontinuous nature of the roll-wave model on which the present stability analysis is based.
}
Presented results are normalised by the VKH growth rates of the respective uniform flows as provided by \eqref{eq:VKH_dispersion_eq}. Normalised time and pulsation are
$T = t \cdot\Re(\omega\_{VKH})$, $\Omega = \omega / \Re(\omega\_{VKH})$, respectively.

The first case, Figure~\ref{fig:1deg:wavelength}, is a one degree upwards-inclined pipe with superficial velocities $\USL= \unitfrac[0.125]ms$ and $\USG = \unitfrac[3.50]ms$ (in the corresponding uniform flow.) Resulting liquid levels are low, with $\ol\al/\area = 0.11$.
Figure~\ref{fig:1deg:wavelength:stab} show the roots of $\Omega$ satisfying \eqref{eq:shock_hat:mom_LR}. 
Roots with a positive real component are unstable. 
According to this figure, unstable wavelengths are predicted to be those shorter than about $40\,d$.
Roots whose disturbance period is one ($\Gm=1$) or two ($\Gm=2$) roll-wave lengths are real, located along the abscissa.
$\Gm$ is made continuous in the plot, though only positive integer values can be realised. 
The unstable roots appear to converge towards some singular point, independent of $\Gm$, at higher wavelengths, the influence of which is unclear.

This predictions of Figure~\ref{fig:1deg:wavelength:stab} seem to agree with the direct simulation of Figure~\ref{fig:1deg:wavelength:Roe}, which shows that the wavelengths quickly arrange themselves between $35$ an $120\,d$ in a slowly decaying, oscillating fashion.
\\

A horizontal pipe is used for the second case, Figure~\ref{fig:0deg:wavelength}. 
Here, $\USL= \unitfrac[0.35]ms$ and $\USG = \unitfrac[1.00]ms$, resulting in $\ol \al/\area = 0.49$.
A typical velocity profile for this case is displayed in Figure~\ref{fig:Biberg_profile:two_phase}.
Figure~\ref{fig:0deg:wavelength:stab} show that the range of unstable $\Omega$ converges towards $\Omega = 0$ (which always exist) with increasing wavelength, and then exists only in the left plane after $\lambda \gtrsim 200\,d$. 
The direct simulation of Figure~\ref{fig:1deg:wavelength:Roe} shows waves ranging from around $180$ an $1200\,d$, oscillating strongly in time.

Examining period and decay of the larger oscillations seen in Figure~\ref{fig:0deg:wavelength:Roe}, one finds %an $\Omega$ range in the order 
$\Omega \sim -0.0015+0.04i$, placing the dominating disturbances close by the ordinate in accordance with the long-wavelength stability predictions seen in Figure~\ref{fig:0deg:wavelength:stab}. Similar observations can be made with Figure~\ref{fig:1deg:wavelength}, though the amplitudes and frequencies are harder to make out.
\\

Let us finally attempt to compare gravity-driven free surface flows and pressure-driven gas-liquid flows. 
The gas phase is removed $(\rho\g=\tau\g=\tau\_i=0)$ and a negative pipe inclination $\theta$ drives a free surface flow.
This free surface flow is uniformly stable if the liquid level and flow rate is identical to that in the previous case, and
increasing the flow rate makes the free surface flow \textit{more} stable as opposed to the previous pressure driven flows. 
The pipe inclination and average liquid fraction are chosen such that the liquid flow rate $\ol \ql$ and VKH growth rate, $\Re(\omega\_{VKH})$, are the same as in the previous case. 
This happens at $\theta = -1.27\degree$, $\USL = 0.35\unitfrac ms$, where $\ol \al/\area = 0.28$. 
Figure~\ref{fig:0deg0rhog:wavelength} shows the stability map and simulated wavelength distribution.
Again, oscillation frequency and decay seem to agree with the stability map. 

\begin{figure}[h!ptb]%
\centering
\begin{subfigure}{\imscale\columnwidth}%
\caption{Snapshots cf. Figure~\ref{fig:0deg:wavelength:Roe}}%
\includegraphics[width=\columnwidth]{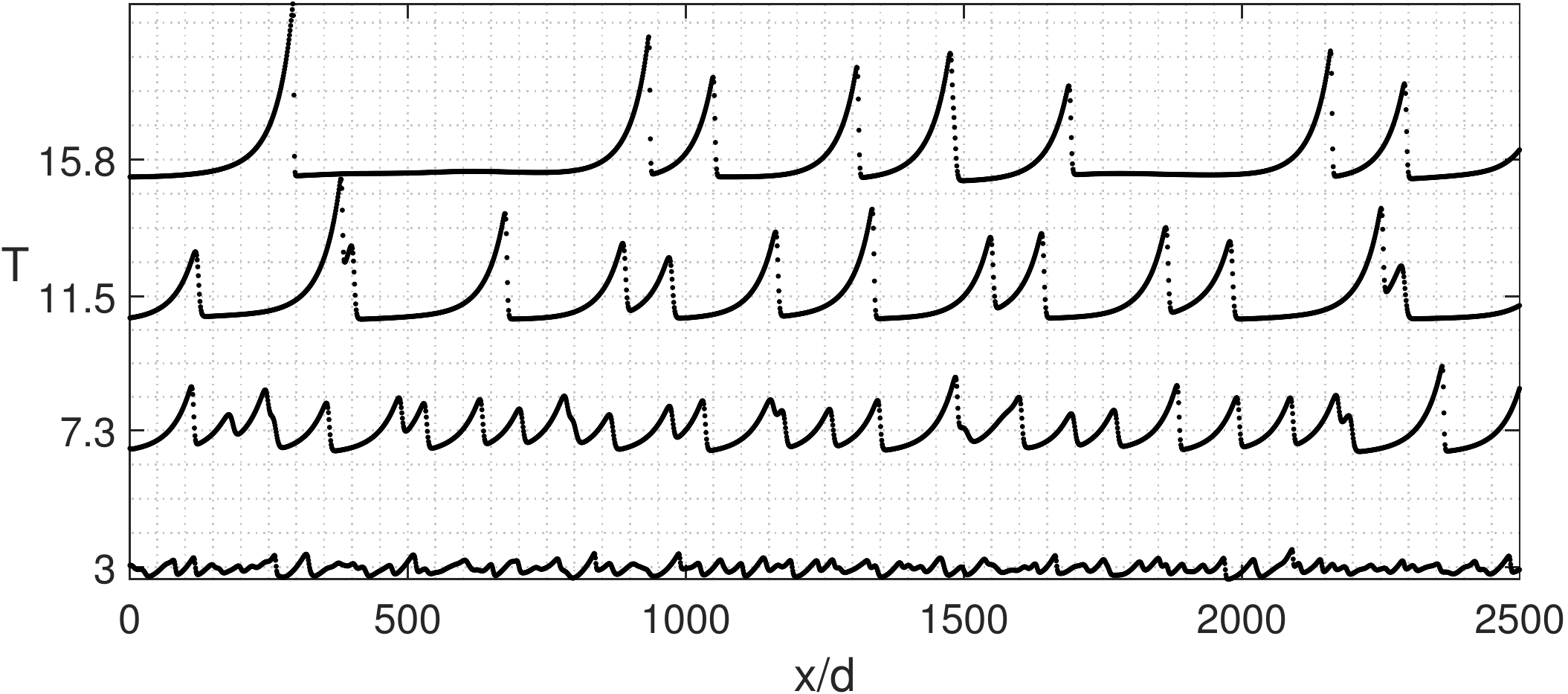}%
\label{fig:snapshots:0deg}%
\end{subfigure}%
\\
\begin{subfigure}{\imscale\columnwidth}%
\caption{Snapshots cf. Figure~\ref{fig:0deg0rhog:wavelength:Roe}}%
\includegraphics[width=\columnwidth]{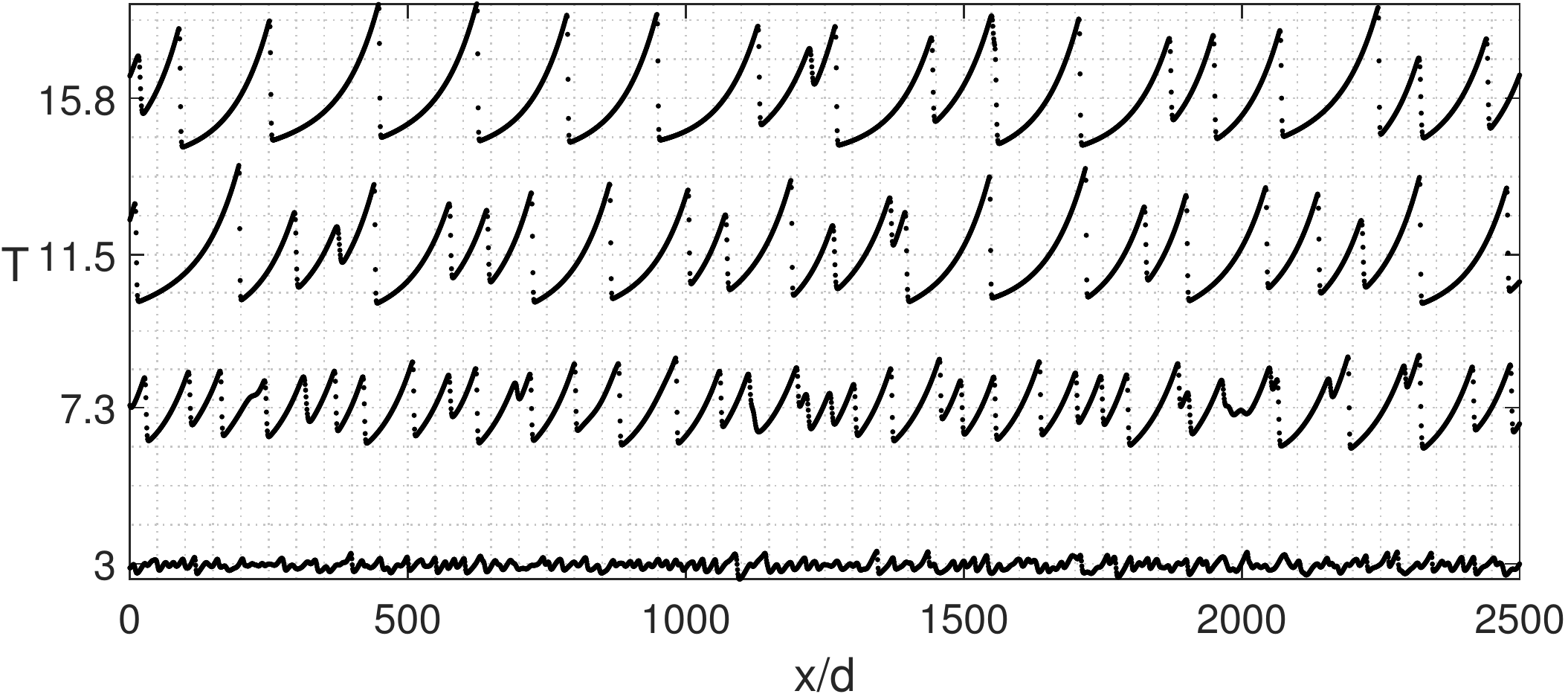}%
\label{fig:snapshots:0deg0rho}%
\end{subfigure}%
\caption{Snapshots of level heights $h$ at specific times early in the simulations. Only a portion of the simulation domains are shown.}%
\label{fig:snapshots}
\end{figure}

Compared to the pressure driven flow of Figure~\ref{fig:0deg:wavelength}, the wavelength distribution of the gravity driven flow develops more steadily, but with some coalescence events occurring at a later stage. 
Figure~\ref{fig:snapshots} shows snapshots of the level heights during the initial developing stages of to two simulations.
The free surface simulation is seen to have wavelengths and wave heights distributed within more narrow bands.
Surviving wavelengths in the free surface flow are of the same order of magnitude as those from the pressure driven flow, and the same can be said for the frequency and decay of the disturbance oscillations. 
The difference in wavelength bands is however not readily evident from comparing the two stability maps of Figure~\ref{fig:0deg:wavelength:stab} and \ref{fig:0deg0rhog:wavelength:stab}. 
The difference can possibly be attributed to the unstable small-wavelength roots, which in the pressure driven flow are seen to be more oscillatory and would thus promote a more irregular wavelength distribution after the initial coalescence stage.

\section{Conclusions}
\label{sec:conclusions}

A stability analysis has been presented which provides predictions in agreement with numerical simulations. 
The analysis seems to give a lower wavelength limit, below which wave trains are unstable. 
Simulations indicate that the wavelength distribution will range close above this lower wavelength limit.

\section*{Acknowledgements}
This work is financed by The Norwegian University of
Science and Technology (NTNU) as a contribution the Multiphase
Flow Assurance programme (FACE.)
%The author would like to thank Tore Fl{\a}tten for his very useful feedback 
%and 
%Kontorbamse for the comforting support.

\appendix

\section{A Roe Scheme}
\label{sec:Roe}
The base model \eqref{eq:base_model:abs} may be written in terms of a Jacobian $\Jac = \pdiff \bff\bw$ as follows:
\begin{equation}
\wt\bw_t + \Jac \wt\bw_x =  \wt\bs,
\label{eq:base_model_Jacform}
\end{equation}
with the Jacobian
\begin{equation}
	\Jac = \frac{1}{\rho\m}
	\begin{pmatrix}
	\br{\rho u}\m & 1\\
	\kappa^2 & \br{\rho u}\m 
	\end{pmatrix},
\label{eq:base_model:J}
\end{equation}
and
\begin{equation*}
	\kappa = \sqrt{\rho\m\my\dHofal - \frac{\rho\l \rho\g}{\al \ag} \br{u\g-u\l}^2}.
\end{equation*} %

Roe's approximate Riemann solver \cite{Roe_original} is among the most popular finite volume schemes for non-linear hyperbolic problems. 
Its main principle lies in linearising \eqref{eq:base_model_Jacform} and solving the Riemann problems
\begin{equation}
\begin{gathered}
 \wt\bw_t + \hat \Jac\of{\bw\_R,\bw\_L }  \wt\bw_x = 0
\\
\wt\bw\of{x,0} = \bw\_L,\: (x<0);\quad \wt\bw\of{x,0} = \bw\_R,\: (x>0)
\end{gathered}
\label{eq:Roe_problem}
\end{equation}
at the cell faces. 
 $\hat \Jac\of{\bw\_R,\bw\_L}$ is the so-called Roe-averaged matrix of $\Jac$.
$\hat \Jac$, $\bw\_L$ and $\bw\_R$ are constants respective to each cell face. 
Roe schemes are effective at discontinuities, but they require the formulation of $\hat \Jac\of{\bw\_L,\bw\_R}$ at the cell faces with the  properties that
\begin{enumerate}[i)]
	\item \label{en:Roe:diag} $\hat \Jac\of{\bw\_L,\bw\_R}$ is diagonalizable with real eigenvalues,
	\item \label{en:Roe:consistant} $\hat \Jac\of{\bw\_L,\bw\_R } \rightarrow \Jac\of{\bw}$ smoothly as $\bw\_L,\bw\_R \rightarrow\bw$, and
	\item \label{en:Roe:main} $\hat \Jac\of{\bw\_L,\bw\_R} \diffx{\bw} = \diffx{\bff}$.
\end{enumerate}
Generally, $\psi\_L = \psi\of{\bw\_L}$ and  $\psi\_R = \psi\of{\bw\_R}$.
The first and second properties are required for hyperbolicity and consistency, respectively. 
The third property ensures, by the Rankine-Hugoniot condition, that single shocks of the linear system \eqref{eq:Roe_problem} are shocks of the non-linear system~\eqref{eq:base_model:abs}.

Consider condition \ref{en:Roe:main}) and the following splitting of the flux function:
\begin{equation}
\bff = \bfu + \bfh
\label{eq:bff_split}
\end{equation}
where
\begin{align*}
\bfu &= \br{\al u\l, \diffk{\rho u/2}}^T,
&
\bfh &= \br{0, \my h}^T.
\end{align*}
We need a suitable integration path over which $\bff$ is easily evaluated; $\bff$ is written in terms of a parameter vector $\bwp$, rendering it  a low-order polynomial. Primitive variable are suitable in the case of $\bfu$, \ie,
\begin{equation*}
\bwp = \br{\al,\ag,u\l,u\g}^T.
\end{equation*}
Note that $\pdiff\bw\bwp$ is constant and that $\pdiff {\bfu}{\bwp}$ is linear in $\bwp$.
A linear path
\[
\bwp=\bwpt\of \ws = \bwp\_L+\diffx\bwp \ws
\]
is chosen for the integration of $\bfu$. We get
\begin{align}
\diffx {\bfu} 
&= \int\_L\^R\! \mr d\bfu 
= \int_0^1\! \pdiff {\bfu}{\bwp}\of{\bwpt\of{\ws}} \pdiff{\bwpt}{\ws} \,\mr d \ws
= \pdiff {\bfu}{\bwp}\of{\ol\bwp} \diffx\bwp \nonumber
\\
&= \pdiff {\bfu}{\bw}\of{\ol\bwp} \pdiff {\bw}{\bwp} \diffx\bwp
= \Jacu\!\of{\ol\bwp} \diffx\bw,
\label{eq:bff_ast}
\end{align}
where $\ol \bwp = \frac12(\bwp\_L+\bwp\_R)$. The fourth expression is a result of $\pdiff {\bfu}{\bwp}$ being linear in $\bwp$, and the fifth and sixth from $\pdiff\bw\bwp$ being constant.
$\Jacu = \pdiff{\bfu}\bw$ is the Jacobian of $\bfu$, equalling \eqref{eq:base_model:J} without the $\rho\m \my\dHofal$ term.

Now consider $\bfh$. We write
\begin{align}
\diffx{\bfh} 
= \br{0, \my \diffx h}^T
= \Jach\of{\a\_{\ell,L},\a\_{\ell,R}}\diffx\bw,
\label{eq:bff_apo}
\end{align}
where
\begin{equation*}
\Jach = 
\begin{pmatrix}
0&0\\ \my {\diffx h}/{\diffx \al} &0
\end{pmatrix}.
\end{equation*}
Inserting \eqref{eq:bff_ast} and \eqref{eq:bff_apo} into \eqref{eq:bff_split},
\begin{equation*}
\diffx \bff = \diffx {\bfu} + \diffx {\bfh} = \br{\Jacu\!\of{\ol\bwp} + \Jach}\diffx\bw,
\end{equation*}
the Roe average matrix
\begin{equation}
\wh \Jac = \Jacu\!\of{\ol\bwp} + \Jach\of{\a\_{\ell,L},\a\_{\ell,R}}
\label{eq:Roe_matrix}
\end{equation}
is seen to be the Jacobian \eqref{eq:base_model:J} constructed from average primitive variables $\ol \ak$ and $\ol {u\k}$, with ${\diffx h}/{\diffx \al}$ replacing $\dHofal$. We use $\dHofal\of {\ol \al} $ close to $\diffx \al = 0$ to avoid numerical 0/0-issues.

Once $\hat \Jac$ is formulated,
\begin{equation}
\bff\of{0,t}= \frac12\br{\bff\_R+\bff\_L} -\frac12 \big|\hat \Jac\big| \diffx{\bw}
\label{eq:Roe_scheme}
\end{equation}
provides the solution of the linearised problem \eqref{eq:Roe_problem}.
Here, 
\begin{equation*}
\big|\hat \Jac\big|= \hat \LL\inv\big|\hat\Lamb\big|\hat\LL
\end{equation*}
the `hat' indicating the Roe intermediate state which in \eqref{eq:Roe_matrix} is the state of arithmetically averaged primitive variables and ${\diffx h}/{\diffx \al}$ replacing $\dHofal$.
Absolute eigenvalue and eigenvector matrices are
\begin{align*}
\big|\Lamb\big| &= 
\begin{pmatrix}
	\big|\lambda^+\big| & 0 \\
	0 & \big|\lambda^-\big|
\end{pmatrix}
&
\text{and}&&
\LL &= 
\begin{pmatrix}
	1 & 1/\kappa \\
	1 & -1/\kappa
\end{pmatrix}
\end{align*}
respectively,
\begin{equation}
	\lambda^\pm = \frac{(\rho u)\m \pm \kappa}{\rho\m}
\label{eq:eigenvalues}
\end{equation}
being the eigenvalues of $\Jac$. 

Integrating \eqref{eq:base_model:abs} in space and time over a control volume cell $j$ yields the common finite volume expression
\begin{equation}
\bw_j\nn = \bw_j - \frac{\dt}{\dx} \big\langle\bff\Jph-\bff\Jmh\big\rangle + \dt \an{\bs_j}
\label{eq:fluxes_fixed_schemes}
\end{equation}
where `new' refers to the state at the next time level, 
the $j$-index to the spatial average of cell $j$
and 
the angle brackets to the temporal average over the time step.
The solution \eqref{eq:Roe_scheme} is here applied directly to each average cell flux $\big\langle\bff\Jmh\big\rangle$ in \eqref{eq:fluxes_fixed_schemes} without spatial reconstruction: $\bw\_R = \bw_j$, $\bw\_L = \bw\Jm$.
Each time step is chosen 
$
\dt = \CFL\,\dx/\max_{j,\pm}\big|\hat\lambda_{j-\frac12}^\pm\big|.
$
The numerical tests presented herein are never in danger of promoting entropy violations in the Roe scheme, which may happen if an expansion fan straddles the time axis of problem~\eqref{eq:Roe_problem}. See \egg~\cite{LeVeque_2002_finite_volume_methods} for entropy corrections.

\bibliographystyle{plainnat}%{plain}% apsrev
{\footnotesize
\bibliography{refs_PhD}}

\end{document}